# Research quality evaluation by AI in the era of Large Language Models: Advantages, disadvantages, and systemic effects

Mike Thelwall, Information School, University of Sheffield, UK.

Artificial Intelligence (AI) technologies like ChatGPT now threaten bibliometrics as the primary generators of research quality indicators. They are already used in at least one research quality evaluation system and evidence suggests that they are used informally by many peer reviewers. Since using bibliometrics to support research evaluation continues to be controversial, this article reviews the corresponding advantages and disadvantages of AI-generated quality scores. From a technical perspective, generative AI based on Large Language Models (LLMs) equals or surpasses bibliometrics in most important dimensions, including accuracy (mostly higher correlations with human scores), and coverage (more fields, more recent years) and may reflect more research quality dimensions. Like bibliometrics, current LLMs do not "measure" research quality, however. On the clearly negative side, LLM biases are currently unknown for research evaluation, and LLM scores are less transparent than citation counts. From a systemic perspective, the key issue is how introducing LLM-based indicators into research evaluation will change the behaviour of researchers. Whilst bibliometrics encourage some authors to target journals with high impact factors or to try to write highly cited work, LLM-based indicators may push them towards writing misleading abstracts and overselling their work in the hope of impressing the AI. Moreover, if AI-generated journal indicators replace impact factors, then this would encourage journals to allow authors to oversell their work in abstracts, threatening the integrity of the academic record.

## Introduction

Post-publication research quality evaluation is a time consuming but important part of modern research systems. On the basis that not all research is equal and the need to hire or reward scholars producing somehow "better" research, an unknown but substantial amount of expert time is devoted to post-publication research quality evaluation. For example, for the UK Research Excellence Framework (REF) 2021, 185,594 research outputs were assessed by 1120 experts (mainly senior professors) over a year, with the results determining UK block research funding grants until 2029. More informally, academic appointments and promotions anywhere in the world may entail experts assessing the work of shortlisted candidates.

Because research quality judgement is a slow task, there has been a natural tendency to look for quick alternatives as quality proxies, such as journal prestige, journal citation rates, or article citation counts. This has proven controversial with strong advocates and opposition. Recently, however, Large Language Models (LLMs) like ChatGPT have started to emerge as an alternative to bibliometrics for supporting research evaluations. This article assesses the potential of LLMs of this task and reviews the issues that may be raised by their apparent new role in research evaluation.

## Expert review and research quality

For important tasks, "research quality" is often judged by peer or expert review, but the concept is not always defined. In systematic examples, guidelines are sometimes created that explicitly or implicitly define quality in a particular context. A comparison of these guidelines revealed that the meaning of research quality varies between contexts. For example, clarity might be considered central to a mathematics journal, and societal impacts might be the sole consideration for a commercial research funder. Nevertheless, most guidelines include rigour, originality, and significance (scientific and/or societal) as important factors for reviewers to consider (Langfeldt et al., 2020). Thus, it seems reasonable to regard these three as the core dimensions of academic research quality, whilst acknowledging that it is not a fixed concept.

Peer review in terms of academics reviewing each other's work submitted to journals, conferences, or book publishers, plays a central gatekeeping quality control purpose as well as providing feedback for improvement. Another type is post-publication peer review, which involves peer review type evaluations of academic research after it has been published, as a type of self-correction function for science (Bordignon, 2020; Da Silva & Dobránszki, 2015; Hunter, 2012; Harms & Credé, 2020; Winker, 2015). In contrast, this article is concerned exclusively with post-publication expert review, however, where the purpose is primarily or exclusively to assess research quality. The result might be a numerical score (e.g., 3/10 or two stars out of 4), perhaps accompanied by a justification, and perhaps with separate scores for each dimension.

Expert or peer review seem to be regarded as best able to assess research quality. Based on the three quality dimensions, disciplinary expertise is needed to judge the rigour of a study (e.g., the suitability of the research design, the appropriateness of the statistics, the exhaustiveness of the argumentation), and its originality (by assessing the extent to which the topic, methods/approach or findings are novel). Judging significance is less obviously an expert task because it involves guessing the future. Nevertheless, an expert might be expected to be more capable than most of estimating the likelihood that research gains traction within a field (e.g., if it is on a hot topic), or has societal value (e.g., in comparison to which work in the field has or hasn't generated societal impact in the past). Overall, however, the fundamental reasons why disciplinary expertise seems necessary for effectively evaluating the quality of academic research are that each academic output is unique and complex, and assessing its rigour, originality, and (likely future) significance seem to require a deep understanding of the field(s) of the output.

A complicating factor is that expert quality judgements depend on the nature of the expertise of the assessor, so different ostensibly similar research evaluation tasks with different types of assessors can assess research differently. If an output in a speciality is assessed by someone from that speciality (e.g., in recruitment for a postdoc position), then they can be expected to give a sharp evaluation of rigour and originality and to assess significance at least partly from the perspective of the speciality. In contrast, if the same output is assessed by a non-specialist from the same field (e.g., a REF assessor), then they would probably be less able to judge rigour and originality, or may judge them from the wider field perspective, and may judge significance primarily from the wider field perspective (e.g., does the output have implications outside of its speciality?) The situation would be different again if the assessor was a non-specialist (e.g., someone on an interdisciplinary panel selecting award recipients) who had no ability to assess rigour and might guess at originality and significance from a general science-wide perspective.

A fundamental problem with peer and expert review is that there is no objective truth in nature (except perhaps in pure maths) and so all science is subjective (Strevens, 2020). Thus, no study can be fully rigorous, and the rigour of an output is therefore a judgment about the extent to which the authors have successfully reduced the chances that their findings are false.

For example, whilst a nineteenth chemistry experiment using spring water for an aqueous solution might be judged rigorous for its time, a similar contemporary experiment might be unacceptable without industrially produced ultrapure water. The problem of subjectivity in science can be compounded by researchers having beliefs, expectations, and moral perspectives. These can lead them to be more critical of, or cautious with, studies that contradict their understanding of the world and therefore be more ready to allocate them lower scores, particularly for rigour. Whilst this can occur at any scale within science, it sometimes translates into the better-known phenomenon of competing paradigms, with adherents being sceptical of claims from competitors (e.g., nature vs. nurture, qualitative vs. quantitative, gene-centric biology vs. systems biology).

An additional problem with peer and expert review is personal bias. Particularly if they lack the skill or time for an effective evaluation, an assessor may consciously or unconsciously be influenced by factors unrelated to the quality of the output, such as the writing style, the gender, nationality, ethnicity, or reputation of the author, or the prestige of their institution. A reviewer may also know the author(s) and be influenced by whether they like them.

Whilst the above discussion is theoretical, reviewers seem to be often uncertain (Barnett et al., 2024), and many empirical studies have confirmed that disagreement between peer reviewers is common (Feliciani et al., 2022; Thelwall & Hołyst, 2023). It is therefore reasonable to believe, albeit with less evidence, that disagreement between expert reviewers for post-publication assessments would also be common.

## Bibliometrics and responsible uses

Bibliometric research quality indicators are typically based on counting citations to academic outputs because citations can reflect influence (Merton, 1973). On this basis, citation counts might be indicators of scholarly significance. In practice, citations can reflect, or be influenced by, many factors other than direct influence (Kousha & Thelwall, 2024; Tahamtan et al., 2016). Despite this complicating factor, evidence that citation-based indicators can weakly or moderately associate with expert research quality judgements in health, physical and life sciences as well as economics (Thelwall et al., 2023b) means that it is technically valid to use them as research quality indicators. Similarly, the citation rate of the publishing journal tends to weakly or moderately associate with expert research quality judgements in the same fields (Thelwall et al., 2023c), so it is also technically valid to use journal citation rates as article quality indicators (Waltman & Traag, 2020).

In the above discussion, the term "indicator" is used in its technical sense of a quantity that associates with research but does not necessarily measure it and is not necessarily accurate. In practice, article-based indicators tend to be field and year normalised and expressed as citation ratios when used in large scale evaluations.

The majority view within the research evaluation community seems to be that peer review is always better than bibliometric indicators for research evaluation but that indicators can play a supporting role. This is the position of the Leiden Manifesto (Hicks et al., 2015), the Metric Tide (Wilsdon et al., 2015), DORA (sfdora.org) and CoARA (coara.eu). One reason for this is that citations might often directly reflect scholarly influence but could only ever be indirect indicators of rigour, originality, and societal significance. In addition, the cited work might be criticised by the citing work, and the cited work might be subsequently discredited. These are valid reasons against using citation-based indicators for individual articles but not for large sets of articles because positive correlations overall with research quality judgements show that these factors tend to cancel out. This does not rule out the possibility of systematic citation biases, however, and these can be serious, depending on the sets of articles compared. For example, if using bibliometric indicators to compare departments in a broad field, then the results would be biased in favour of departments working in high citation specialties. Thus, caution is always needed with bibliometric indicators.

Some within the research evaluation community regard citations rather than peer review as the best evidence of research quality (Rushforth & Hammarfelt, 2023). They can point to evidence of subjectivity and biases within peer review and claim that citation counts constitute the cumulative decisions of many expert authors in the field. This does not address the fact that citations only directly reflect the scholarly impact dimension of research quality, however.

A third perspective within the wider research evaluation community, and expressed in the Leiden Manifesto (Hicks et al., 2015) and the Metric Tide (Wilsdon et al., 2015), is the need to consider the systemic effects of any evaluation process (Rushforth & Hammarfelt, 2023). The use of citation-based indicators in important research evaluations can incentivise researchers to chase citations instead of research quality. For example, this may drive them towards generating scientific impacts (e.g., by introducing new research methods) rather than societal impacts (e.g., by assessing the value of a method to an end user community) and drive scholars towards theoretical rather than applied research. DORA (sfdora.org) seems to be a response to the overvaluing of individual journals by scholars, particularly in the biomedical community, which may have pushed some academics too far towards conducting the type of research that can be published in high impact journals rather than the type of research that would be more widely valuable. Expert review seems to have a substantial advantage from the systemic perspective because, even if the experts are poor at evaluating research quality, expert review only incentivises research adhering to the statement of quality used in the evaluation process. A partial exception to this is that if the evaluator identities are known far enough in advance, then the people evaluated might be incentivised to tailor their outputs to the type that likely assessors are perceived to value.

## Machine learning, LLMs, and Generative AI

Machine learning is a generic term for artificial intelligence programs that learn to complete a task from examples of it. It is particularly useful when there are multiple factors that are known to influence a target variable (e.g., research quality) but the relationship between them is complex or non-linear. Machine learning has been used to predict long term citation counts based on early citation counts, journal impact factors, authorship team size and other variables (Qiu & Han, 2024) or through LLM text analysis (Zhao et al., 2024). Machine learning long term citation predictions do not ever seem to have been used in practice for research evaluations, perhaps because they reduce transparency and add complexity. There has also been an attempt to predict article *quality* scores with machine learning based on bibliometric inputs (e.g., article and journal normalised citation rates, author numbers, title and abstract words and short phrases), producing the most accurate results in the health, life and physical sciences and economics, but never exceeding 75% accurate on the four-point scale used (Thelwall et al., 2023a). The target four-point quality score was that used in the UK REF2021: internationally relevant 1*, internationally relevant 2*, internationally excellent 3*, and world leading 4*, all in terms of originality, rigour and significance. This solution would need to be integrated with large scale expert reviews, however, because it needs a large amount of training data to be effective. This makes it of niche value, only useful to aid large scale national research evaluations.

Large Language Models are technically a form of machine learning because they are essentially generic network structures that learn about language through being fed huge volumes of text. Nevertheless, they are usefully considered as a separate AI category. There are two main types of LLM, discriminative and generative. Whilst discriminative LLMs like BERT (Bidirectional Encoder Representations from Transformers) (Devlin, 2018) can be fed with text and report context about it, generative AI based on LLMs, including ChatGPT, can write new text. Early generative AI LLMs worked by predicting the likely next tokens in an input sequence, sometimes with a probability factor for randomisation. For example, feeding an early LLM with "my tea is too" would be likely to predict/generate "hot" or "cold" but not "blue" or "is" as the next word. With huge amounts of input text, entire coherent sentences and paragraphs can be reliably produced. ChatGPT took this further by training LLM systems not just on general text but also on question-response data (and probably other tasks like software code writing) in a process known as reinforcement learning from human feedback (Ouyang et al., 2022). This makes the system more effective at responding to a wide range of user input requests. It harnesses real user data from the web version to improve its performance continually or periodically.

At the time of writing (early 2025), the set of public Generative Pre-trained Transformers (GPTs) generative AI LLMs had expanded to include Google Gemini, Claude, Facebook's LLaMA (sometimes written Llama, available at meta.ai), and DeepSeek, as well as derivatives like Microsoft Copilot (based on ChatGPT-4).

In addition to the big public GPTs, there is a range of open source shared GPT models that can be used to perform the same function online, many of which are available at Huggingface.co. These include free versions of commercial models, like DeepSeek and LLaMA (e.g., Masalkhi et al., 2024), as well as models created by researchers, such as OpenGPT-2 (Cohen & Gokaslan, 2020). The same site also shares non-GPT LLMs. For example, the many Bidirectional Encoder Representations from Transformers (BERT) variants do not generate text but can be used to classify text (e.g., for sentiment) (Sun et al., 2019). Although offline LLMs are less powerful, they are free and can be safely used (including in document processing pipelines) with private data. Automated processing is also possible with the online commercial LLMs through their Applications Programming Interfaces (APIs), allowing batch processing of documents or requests.

## LLM-generated research quality indicators

LLM-based generative AI systems like ChatGPT can be directly tasked with research quality evaluation by feeding them with a research quality definition and an article and then asking them to score that article. This is very different from bibliometrics and traditional machine learning because it mimics expert review and focuses on the article text rather than harnessing citations and/or metadata. Many studies have now shown that LLMs can give useful peer review style feedback on academic papers (Lu et al., 2024; Du et al., 2024; Zhou et al., 2024) and can make predictions about acceptance/rejection that correlate positively with editorial decisions (Zhou et al., 2024; Thelwall & Yaghi, 2024a; Zhuang et al., 2025). Early research with ChatGPT 4o has now also given promising results for research quality evaluation too. Of course, some reviewers also harness ChatGPT to help with writing (Liang et al., 2024).

Several small-scale and medium-scale studies and one large scale study have found positive correlations between expert quality scores or peer review decisions and ChatGPT's predictions. A study of 21 medical papers found a statistically significant correlation between human and ChatGPT 3.5 peer review recommendations (accept, accept with revisions, reject) but not for ChatGPT 4o (Saad et al., 2024), possibly due to the small sample size in the latter case. The remaining three studies have all used the REF quality scoring guidelines and scoring system (1*, 2*, 3* or 4*) (REF, 2019). The first uploaded 51 PDFs and Word documents of published or unpublished journal articles from the field of library and information science to the web interface of ChatGPT 4o, asking for a REF score for each one. The results correlated positively but weakly (0.20) with scores given by the author, but the correlation increased to moderate (0.51) when the articles were submitted 15 times and the results averaged (Thelwall, 2024). A follow-up study with the same set of papers used the ChatGPT API instead of the web interface and only submitted the titles and abstracts, without the full texts. This achieved even higher correlations (0.67) between the ChatGPT average (over 30 iterations) and the author's score. The results were worse (lower correlations) with the article title alone so the main power of ChatGPT seems to be in interpreting the author's information about (or claims) for originality, significance and/or rigour from their abstract (Thelwall, 2025a). Similar but slightly weaker positive results have also been found for Google Gemini 1.5 Flash, but PDF inputs produced stronger correlations than titles and abstracts in this case (Thelwall, 2025b).

Medium scale studies have tended to replicate the small-scale study results for individual fields. In the REF2021 category of Clinical Medicine, ChatGPT 4o-mini scores for 9872 articles found a weak positive correlation with the average research quality of the department associated with the article. Average ChatGPT scores correlated more positively with average departmental scores for articles at the journal level, although the results suggested that the dry reporting style of some prestigious medical journals tended to undermine their articles' scores (Thelwall et al., 2024). For the related goal of novelty scoring, LLMs have been shown to have some ability to predict novelty scores for computer science papers from one conference from the introduction, results, and discussion but not from the full text (Wu et al., 2025; see also: de Winter, 2024). For a different output type, ChatGPT 4o-mini scores based on titles and blurbs/abstracts weakly but positively associate with book citation rates in the social sciences and humanities (Thelwall & Cox, 2025).

The one large-scale analysis of ChatGPT quality scores so far selected articles from high and low scoring departments in each of the 34 Units of Assessment (UoAs – these are essentially broad fields or sets of similar small fields) and correlated the ChatGPT 4o quality scores (averaged over 30 iterations; titles and abstracts submitted through the API) with the departmental average scores. The results were positive in all UoAs except Clinical Medicine (-0.15) and were weak (0.05 to 0.3) mainly in the arts and humanities, moderate (0.3 to 0.5) mainly in the social sciences and strong (0.5 to 0.8) mainly in the health and physical sciences and engineering (Thelwall & Yaghi, 2024b). Although the results were promising, they only used UK articles and, most importantly, relied on public data in terms of the departmental average REF scores. Thus, whilst they are consistent with ChatGPT having a near-universal ability to detect research quality, they do not prove it, because ChatGPT might have leveraged public information about departmental REF quality profiles when scoring individual articles. The reason for the outlying field, clinical medicine, might be that abstracts in clinical areas seem to report facts free of interpretation in their abstracts, giving ChatGPT less context to deduce significance. For completeness, there is also an even larger-scale investigation with ChatGPT 4o-mini average quality scores for 90% of REF2021 journal articles. It showed that ChatGPT scores tended to positively correlate more highly with raw citation counts than with field (and year) normalised citation rates (Thelwall & Jiang, 2025).

From a different perspective, another large-scale study has investigated the potential biases of LLMs for research evaluation, using 117,650 articles from five years (2003, 2008, 2013, 2018 and 2023) in 26 Scopus fields. It found a slight tendency for more recent articles to receive higher scores over the 20 years. There were also field differences in average ChatGPT scores, and articles with longer abstracts tended to receive higher scores (Thelwall & Kurt, 2024). These disparities are not necessarily biases, however, because of the lack of a ground truth in the study. For example, it is possible that research quality has improved over time or that for some reason better quality articles tend to have longer abstracts.

Whilst the results above are not conclusive, they are suggestive of, and consistent with, LLM-based quality evaluations being already superior to bibliometrics as research quality indicators, although with clear and not yet well understood biases. Given that LLM technologies are still evolving, it seems reasonable to consider potential use within practical research evaluation contexts.

### Advantages

LLM-based quality evaluations seem to have at least three clear advantages over bibliometrics for research quality indicators. These are summarised here.

**More accurate**: ChatGPT 4o scores seem to be more accurate than bibliometrics in the sense of higher correlations with human scores in most fields (Thelwall & Yaghi, 2024b). Other factors being equal (e.g., biases) they are therefore more useful in any task that bibliometrics are currently used for and in theory could be used for purposes that bibliometrics are currently not.

**Greater coverage of science (years, fields)**: Other than those for journals, citation-based indicators need at least two or three years to mature and so are not applicable to recently published research. In contrast, LLMs can be applied to research of any age, as far as is currently known. This is a substantial advantage given that the most recent research seems likely to be the most relevant to evaluate in practical contexts. In addition, citation-based indicators are useless (very weak or zero correlations with expert quality scores) for all arts and humanities and many social sciences (Thelwall et al., 2023b), whereas ChatGPT 4o seems only to be useless for clinical medicine.

**More quality dimensions assessed**: Whilst citation-based indicators only directly reflect scholarly impact and are at best indirect indicators of rigour, originality, and societal impact, LLM research quality ratings can, in theory, cover all dimensions and there is some, albeit weak (Thelwall & Yaghi, 2024a), evidence that it can in some contexts. Whilst LLMs rely on the dimensions of quality defined for them in the systems instructions, and these have human limitations in terms of potential inaccuracies and biases, humans are the ultimate gatekeepers of research quality and so there does not seem to be an alternative.

### Similarities

An important similarity with bibliometrics is that LLMs do not "measure" research quality. In the case of citation-based indicators, they primarily reflect one type of scholarly impact and so are not an overall research quality measure. Whilst there has been a systematic attempt to ameliorate this issue with alternative quantitative indicators, altmetrics, these have not filled the gap in most contexts. LLMs currently work most effectively with article titles and abstracts and therefore are clearly guessing at research quality rather than measuring it. Moreover, they may ignore or undervalue less common types of research or unusual contributions, including culturally specific work that differs from the mainstream despite being of equal quality. Whilst an LLM-based evaluation might use full texts rather than titles and abstracts, this would not mean that research quality was being measured. This is because more accurate results with titles and abstracts alone would be consistent with full-text-based evaluations primarily assessing titles and abstracts, with the rest of the text primarily confusing the LLM.

### *Disadvantages*

There are clear disadvantages of LLMs compared to bibliometrics, as of September 2024. Some of these may lessen over time.

**Unknown biases**: AI systems can learn biases from their training data and even generate new biases because of how their algorithms work (Kordzadeh & Ghasemaghaei, 2022). Thus, it is plausible that LLMs have learned relevant human biases, such as gender prejudice or favouritism towards the work of successful authors or those from prestigious institutions. They may also have learned a hierarchy of methodological approaches, disciplines or research topics that would influence their quality judgements. Despite the current absence of evidence that these or any other biases exist (with the possible exception of document age, field, and abstract length: Thelwall & Kurt, 2024), it is not reasonable to assume that they don't and therefore applications should be cautious and new research into bias is needed. It is possible that identified AI biases could be mitigated, for example by identifying and correcting for particular dimensions, or by emphasising in prompts the need to be unbiased, but this is an unexplored area. In comparison, bibliometric data has been tested for many types of bias. There seem to be minor gender bias at the article level (Thelwall, 2020) but more substantial career biases (Kelly & Jennions, 2006) for citations as well as perhaps national self-citation biases and national biases due to limited coverage of a nation's publications, causing lost citations (Pendlebury, 2020). These are known factors, and the biases overall are probably most important for international comparisons. Any responsible application of either bibliometrics or LLMs should at least consider the possibility of biases and take steps to mitigate them and/or decide whether biases are too serious for the data to be used.

**Lower transparency**: Bibliometric data is relatively transparent, although there are some opaque or obscure elements. These include the decision-making process for including documents to index in the citation database (which can include human factors), and the details of the algorithms for extracting references from documents and matching them with the cited works. In addition, the thought process of each citer is unknown. In contrast, LLMs are almost fully opaque. Although the LLM architecture is known (transformer variants), the data used to train commercial variants is unknown, as are the workings of the many additional algorithms used for each complete system. Most significantly, however, the algorithms are so complex that even full knowledge of them would give little insight into how they can successfully score articles for quality or the conditions under which they would make mistakes. Although each LLM score can be accompanied by a detailed explanation and justification, these seem to be too vague to be helpful (Thelwall, 2024, 2025a).

**Lack of research into different use contexts**: Another disadvantage of LLMs is that research into their value for research quality assessment is limited now (February, 2025) and it would be helpful to have a range of different studies and approaches to confirm or contradict those published so far. This is needed to give end users confidence in the value of the LLM scores. Moreover, since LLMs have been only used once so far in research evaluations (for project grants: Carbonell Cortés, 2024), there is a lack of experience about how to use them and of the approaches that would be most effective.

**More serious side effects of gaming**: An important disadvantage of LLM scores is their potential for gaming. Whilst citations can be manipulated through excessive author self-citations (or journal self-citations for journal citation rate indicators), citation cartels, or other strategic citation practices (e.g., Baccini et al., 2019) this does not seem to have been a major concern in any research evaluation so far, other than for journal rankings (e.g., Caon, 2017). In contrast, the potential of LLMs to be gamed is unknown and therefore cannot be assumed to be negligible. For instance, if LLMs primarily leverage author claims in abstracts then the use of LLMs in serious research evaluation tasks would incentivise authors to overclaim as much as they could get away with. Reviewers and journal editors would then need to control any perceived exaggeration. If LLM-based indicators (e.g., average scores) were used for journals then this would give a similar incentive to editors and publishers to allow inflated author abstract claims. In either case, this would undermine the usefulness of abstracts as information carriers and the integrity of research. It might also reduce the effectiveness of LLMs.

## Conclusion: Responsible uses of LLMs in research evaluation

As the above discussion suggests, LLM scores have the technical potential to complement or surpass bibliometrics for research quality indicators, but there are too many unknowns yet to use them in any important context. If more information can be gained about their biases, limitations, and potential for gaming then it might be possible to start using them in minor roles to support peer review. In the longer term, successful applications, and a lack of issues from gaming might allow them to be used in less minor roles, taking over from bibliometrics. For example, they might be offered instead of bibliometrics as supporting information for expert reviewers in future versions of the UK REF national research evaluation exercise.

## Declarations

**Funding and/or Conflicts of interests/Competing interests**: The first author is a member of the Distinguished Reviewers Board of this journal and was funded by an ESRC Metascience grant.

## References

Baccini, A., De Nicolao, G., & Petrovich, E. (2019). Citation gaming induced by bibliometric evaluation: A country-level comparative analysis. PLoS One, 14(9), e0221212.

Barnett, A., Allen, L., Aldcroft, A., Lash, T. L., & McCreanor, V. (2024). Examining uncertainty in journal peer reviewers' recommendations: a cross-sectional study. Royal Society Open Science, 11(9), 240612.

Bordignon, F. (2020). Self-correction of science: a comparative study of negative citations and post-publication peer review. Scientometrics, 124, 1225-1239. https://doi.org/10.1007/s11192-020-03536-z.

Caon, M. (2017). Gaming the impact factor: where who cites what, whom and when. Australasian Physical & Engineering Sciences in Medicine, 40, 273-276.

Carbonell Cortés, C. (2024). AI-assisted pre-screening of biomedical research proposals: ethical considerations and the pilot case of "la Caixa" Foundation. https://www.youtube.com/watch?v=O2DcXzEtCmg

Cohen, V., & Gokaslan, A. (2020). OpenGPT-2: Open language models and implications of generated text. XRDS: Crossroads, The ACM Magazine for Students, 27(1), 26-30.

Da Silva, J., & Dobránszki, J. (2015). Problems with Traditional Science Publishing and Finding a Wider Niche for Post-Publication Peer Review. Accountability in Research, 22, 22 - 40. https://doi.org/10.1080/08989621.2014.899909.

Devlin, J. (2018). BERT: Pre-training of deep bidirectional transformers for language understanding. arXiv preprint arXiv:1810.04805.

de Winter, J. (2024). Can ChatGPT be used to predict citation counts, readership, and social media interaction? An exploration among 2222 scientific abstracts. Scientometrics, 129(4), 2469-2487.

Du, J., Wang, Y., Zhao, W., Deng, Z., Liu, S., Lou, R., & Yin, W. (2024). LLMs assist NLP researchers: Critique paper (meta-) reviewing. arXiv preprint arXiv:2406.16253.

Feliciani, T., Luo, J., & Shankar, K. (2022). Peer reviewer topic choice and its impact on interrater reliability: A mixed-method study. Quantitative Science Studies, 3(3), 832-856.

Harms, P., & Credé, M. (2020). Bringing the review process into the 21st century: Post-publication peer review. Industrial and Organizational Psychology, 13, 51-53. https://doi.org/10.1017/iop.2020.13.

Hicks, D., Wouters, P., Waltman, L., De Rijcke, S., & Rafols, I. (2015). Bibliometrics: the Leiden Manifesto for research metrics. Nature, 520(7548), 429-431.

Hunter, J. (2012). Post-Publication Peer Review: Opening up scientific conversation. Frontiers in Computational Neuroscience, 6. https://doi.org/10.3389/fncom.2012.00063.

Kelly, C. D., & Jennions, M. D. (2006). The h index and career assessment by numbers. Trends in Ecology & Evolution, 21(4), 167-170.

Kordzadeh, N., & Ghasemaghaei, M. (2022). Algorithmic bias: review, synthesis, and future research directions. European Journal of Information Systems, 31(3), 388-409.

Kousha, K., & Thelwall, M. (2024). Factors associating with or predicting more cited or higher quality journal articles: An Annual Review of Information Science and Technology (ARIST) paper. Journal of the Association for Information Science and Technology, 75(3), 215-244.

Langfeldt, L., Nedeva, M., Sörlin, S., & Thomas, D. A. (2020). Co-existing notions of research quality: A framework to study context-specific understandings of good research. Minerva, 58(1), 115-137.

Liang, W., Izzo, Z., Zhang, Y., Lepp, H., Cao, H., Zhao, X., & Zou, J. Y. (2024). Monitoring ai-modified content at scale: A case study on the impact of ChatGPT on AI conference peer reviews. arXiv preprint arXiv:2403.07183.

Lu, Y., Xu, S., Zhang, Y., Kong, Y., & Schoenebeck, G. (2024). Eliciting Informative Text Evaluations with Large Language Models. arXiv preprint arXiv:2405.15077.

Masalkhi, M., Ong, J., Waisberg, E., Zaman, N., Sarker, P., Lee, A. G., & Tavakkoli, A. (2024). A side-by-side evaluation of Llama 2 by meta with ChatGPT and its application in ophthalmology. Eye, 38(10), 1789-1792.

Merton, R. K. (1973). The sociology of science: Theoretical and empirical investigations. Chicago: University of Chicago.

Ouyang, L., Wu, J., Jiang, X., Almeida, D., Wainwright, C., Mishkin, P., & Lowe, R. (2022). Training language models to follow instructions with human feedback. Advances in neural information processing systems, 35, 27730-27744.

Pendlebury, D. A. (2020). When the data don't mean what they say: Japan's comparative underperformance in citation impact. Evaluative Informetrics: The Art of Metrics-Based Research Assessment: Festschrift in Honour of Henk F. Moed, 115-143.

Qiu, J., & Han, X. (2024). An early evaluation of the long-term influence of academic papers based on machine learning algorithms. IEEE Access, 12, 41773-41786.

REF (2019). Panel criteria and working methods (2019/02). https://2021.ref.ac.uk/publications-and-reports/panel-criteria-and-working-methods-201902/index.html

Rushforth, A., & Hammarfelt, B. (2023). The rise of responsible metrics as a professional reform movement: A collective action frames account. Quantitative Science Studies, 4(4), 879-897.

Saad, A., Jenko, N., Ariyaratne, S., Birch, N., Iyengar, K. P., Davies, A. M., Vaishya, R., & Botchu, R. (2024). Exploring the potential of ChatGPT in the peer review process: An observational study. Diabetes & Metabolic Syndrome: Clinical Research & Reviews, 18(2), 102946. https://doi.org/10.1016/j.dsx.2024.102946

Sun, C., Qiu, X., Xu, Y., & Huang, X. (2019). How to fine-tune BERT for text classification? In China national conference on Chinese computational linguistics (pp. 194-206). Cham: Springer International Publishing.

Tahamtan, I., Safipour Afshar, A., & Ahamdzadeh, K. (2016). Factors affecting number of citations: a comprehensive review of the literature. Scientometrics, 107, 1195-1225.

Thelwall, M. (2020). Female citation impact superiority 1996–2018 in six out of seven English-speaking nations. *Journal of the Association for Information Science and Technology*, 71(8), 979-990.

Thelwall, M. (2024). Can ChatGPT evaluate research quality? *Journal of Data and Information Science*, 9(2), 1–21. https://doi.org/10.2478/jdis-2024-0013

Thelwall, M. (2025a). Evaluating research quality with large language models: an analysis of ChatGPT's effectiveness with different settings and inputs. *Journal of Data and Information Science*, 10(1), 7-25. https://doi.org/10.2478/jdis-2025-0011

Thelwall, M. (2025b). Is Google Gemini better than ChatGPT at evaluating research quality? Journal of Data and Information Science, 10(1), 1–5. https://doi.org/10.2478/jdis-2025-0014

Thelwall, M. & Cox, A. (2025). Estimating the quality of academic books from their descriptions with ChatGPT. Journal of Academic Librarianship. https://arxiv.org/abs/2502.08171

Thelwall, M., & Hołyst, J. A. (2023). Can journal reviewers dependably assess rigour, significance, and originality in theoretical papers? Evidence from physics. *Research Evaluation*, 32(2), 526-542.

Thelwall, M., & Jiang, X. (2025). Is OpenAlex Suitable for Research Quality Evaluation and Which Citation Indicator is Best? arXiv preprint arXiv:2502.18427.

Thelwall, M., Jiang, X., & Bath, P. A. (2024). Evaluating the quality of published medical research with ChatGPT. arXiv preprint arXiv:2411.01952.

Thelwall, M., Kousha, K., Wilson, P. Makita, M., Abdoli, M., Stuart, E., Levitt, J., Knoth, P., & Cancellieri, M. (2023a). Predicting article quality scores with machine learning: The UK Research Excellence Framework. *Quantitative Science Studies*, 4(2), 547–573. https://doi.org/10.1162/qss_a_00258

Thelwall, M., Kousha, K., Stuart, E., Makita, M., Abdoli, M., Wilson, P., & Levitt, J. (2023b). In which fields are citations indicators of research quality? Journal of the Association for Information Science and Technology, 74(8), 941-953.

12
Thelwall, M., Kousha, K., Makita, M., Abdoli, M., Stuart, E., Wilson, P., & Levitt, J. (2023c). In which fields do higher impact journals publish higher quality articles? Scientometrics, 128(7), 3915-3933.

Thelwall, M., & Yaghi, A. (2024a). Evaluating the predictive capacity of ChatGPT for academic peer review outcomes across multiple platforms. Submitted.

Thelwall, M., & Yaghi, A. (2024b). In which fields can ChatGPT detect journal article quality? An evaluation of REF2021 results. https://arxiv.org/abs/2409.16695

Thelwall, M., & Kurt, Z. (2024). Research evaluation with ChatGPT: Is it age, country, length, or field biased? arXiv preprint arXiv:2411.09768.

Strevens, M. (2020). *The knowledge machine: How an unreasonable idea created modern science*. Penguin UK.

Waltman, L., & Traag, V. A. (2020). Use of the journal impact factor for assessing individual articles: Statistically flawed or not? F1000Research, 9.

Wilsdon, J., Allen, L., Belfiore, E., Campbell, P., Curry, S., Hill, S., & Johnson, B. (2015). The metric tide: Independent review of the role of metrics in research assessment and management. https://www.ukri.org/publications/review-of-metrics-in-research-assessment-and-management/

Winker, M. (2015). The promise of post‐publication peer review: how do we get there from here? Learned Publishing, 28(2), 143-145. https://doi.org/10.1087/20150209.

Wu, W., Zhang, C., Bao, T., & Zhao, Y. (2025). SC4ANM: Identifying optimal section combinations for automated novelty prediction in academic papers. Expert Systems with Applications, 126778.

Zhao, P., Xing, Q., Dou, K., Tian, J., Tai, Y., Yang, J., & Li, X. (2024). From Words to Worth: Newborn Article Impact Prediction with LLM. arXiv preprint arXiv:2408.03934.

Zhou, R., Chen, L., & Yu, K. (2024). Is LLM a Reliable Reviewer? A Comprehensive Evaluation of LLM on Automatic Paper Reviewing Tasks. In *Proceedings of the 2024 Joint International Conference on Computational Linguistics, Language Resources and Evaluation (LREC-COLING 2024)* (pp. 9340-9351).

Zhuang, Z., Chen, J., Xu, H., Jiang, Y., & Lin, J. (2025). Large language models for automated scholarly paper review: A survey. arXiv preprint arXiv:2501.10326.